\def\spose#1{\hbox to 0pt{#1\hss}}

\def\multleft#1{\hbox to size{\vbox {\halign {\lft{##}\cr #1}}\hfill}\par}
\def\multright#1{\hbox to size{\vbox {\halign {\rt{##}\cr #1}}\hfill}\par}

\def\today{\ifcase\month\or January\or February\or March\or April\or May\or
      June\or July\or August\or September\or October\or November\or December\fi
      \space\number\day, \number\year}
\def\s{\hbox{\phantom{5}}}	

\def\mJy{{\rm\thinspace mJy}}
\def\microJy{{\rm\thinspace $\mu$Jy}}

\def\km{{\rm\thinspace km}}

\def\Mpc{{\rm\thinspace Mpc}}

\def\s{{\rm\thinspace s}}

\def\kmps{\hbox{$\km\s^{-1}\,$}}

\def\kmpspMpc{\hbox{$\kmps\Mpc^{-1}$}}

\def\H2{\hbox{H$_{2}$}}

\voffset- .65in

\documentclass[usegraphicx]{mn2e}

\usepackage{times}
\usepackage{amssymb}
\include{defn}

\begin{document}
\hsize=6truein

\title{The clustering of sub-mJy radio sources in The Bootes Deep Field}
\author[R.J.~Wilman et al.]
{\parbox[]{6.in} {R.J.~Wilman, H.J.A.~R\"{o}ttgering, R.A.~Overzier, and M.J.~Jarvis \\ \\
\footnotesize
Sterrewacht Leiden, Postbus 9513, 2300 RA Leiden, The Netherlands. \\}}
\maketitle

\begin{abstract}
We measure the angular clustering of $\sim 2000$ radio sources in The Bootes 
Deep Field, covering 5.3~deg$^{2}$ down to $S_{\rm{1.4 GHz}}=0.2$\mJy. With 
reference to work by Blake \& Wall, we show that the size distribution of 
multi-component radio galaxies dominates the overall clustering signal, and 
that its amplitude extrapolates smoothly from their measurements above 5\mJy. 
The upper limits on any true galaxy-galaxy clustering are consistent with the 
clustering of sub-\mJy~radio-loud AGN being effectively diluted by the more 
weakly-clustered {\em IRAS}-type starburst galaxies. Source count models imply 
that the survey contains $\simeq 400$ of the latter galaxies above 0.2\mJy~out 
to $z \sim 1 - 2$. Measurement of their clustering must await their 
identification via the optical and infrared data due on this field.
\end{abstract}


\section{INTRODUCTION}
Measurements of the clustering of galaxies can shed light on the formation of 
large scale structure and of the galaxies within it. In the local Universe 
($z < 0.1$), optically-selected galaxies are unbiased tracers of the mass, 
following a spatial correlation function of the form 
$\xi(r)=(r/r_{\rm{0}})^{-\gamma}$, with $r_{\rm{0}}=5.4 h^{-1}$\Mpc~and 
$\gamma=1.8$ (Davis \& Peebles~1983). Results from the 2dF QSO redshift survey 
show that at $\bar{z}=1.3$, QSOs are comparably clustered 
($r_{\rm{0}}=5.7 h^{-1}$\Mpc; Shanks et al.~2001). In contrast, the local 
population of {\em IRAS} starburst galaxies are less strongly clustered 
($r_{\rm{0}} \simeq 3.8 h^{-1}$\Mpc; Saunders et al.~1992), whilst the 
opposite is true of early-type galaxies (for local $L \gtrsim L^{\star}$ 
ellipticals $r_{\rm{0}}=9-11h^{-1}$\Mpc, e.g. Guzzo et al.~1997, Willmer et 
al.~1998). 

Extending such work to higher redshift populations can constrain the galaxy 
systems into which they evolve: e.g. Adelberger (2000) measured the clustering 
of $z \sim 1$ Balmer-break galaxies and derived $r_{\rm{0}}=3.0 h^{-1}$\Mpc~
(comoving). This is consistent with the $z \sim 1$ Balmer-break galaxies being 
both unbiased tracers of mass and the progenitors of local $L^{\star}$ 
galaxies. The comparable comoving space densities of the two populations 
supports this interpretation. In contrast, at $z \sim 3$ the Lyman-break 
galaxies are more strongly clustered (see Porciani \& Giavalisco~2002), 
implying that they are strongly biased relative to the mass distribution. 
Similarly, analysis of the extremely red objects (EROs) demonstrates that 
passively evolving early-type galaxies at $z \sim 1.2$ have 
$r_{\rm{0}}=12 \pm 3 h^{-1}$\Mpc, comparable to that of local $L^{\star}$ 
ellipticals (Daddi et al.~2001; McCarthy et al.~2001). This rules out a 
scenario in which Lyman-break galaxies evolve into local bright ellipticals, 
and requires that the bias evolves with redshift such that the comoving 
density of ellipticals is invariant to $z \sim 1$ and beyond (see also 
Moustakas \& Somerville~2002). 

In contrast, radio surveys probe galaxy clustering over larger scales, as they 
cover large areas of sky with a broad redshift distribution. This does, 
however, reduce the clustering signal when projected on the sky. Nevertheless, 
numerous positive measurements of the angular clustering of radio sources have 
been made. The most complete analysis to date is that of Blake \& Wall~(2002) 
(hereafter BW02), who measured the angular correlation function of sources in 
the 1.4~GHz NVSS from flux thresholds of 50 down to 5\mJy~(see also Overzier 
et al.~2002). They found that the clustering at small angles is due to the 
size distribution of multi-component radio galaxies, and that the 
galaxy-galaxy correlation function has the universal slope of $\gamma=1.8$, 
and  (under certain assumptions) a correlation length of 
$r_{\rm{0}} \sim 6 h^{-1}$\Mpc.


Below $\sim 1$\mJy~(at 1.4~GHz) there is evidence for the appearance of a 
different radio source population which may replace the radio-loud AGN as the 
dominant population below a few 100$\mu$Jy (Windhorst~1984, Windhorst et al.~
1985, Benn et al.~1993). These sources are more distant analogues of the dusty 
starburst galaxies selected by IRAS, and models of the radio source counts 
require that this population undergo substantial luminosity evolution out to 
$z \sim 1$ (Rowan-Robinson et al.~1993; hereafter RR93). Measurements of the 
clustering of the sub-mJy population could thus extend the local IRAS 
clustering measurements of starburst galaxies out to $z \sim 1$, for 
comparison with those of the Balmer-break galaxies, the EROs referred to above 
and the dusty, star-forming EROs at $z \sim 1$ which have $r_{\rm{0}}$ no 
larger than $2.5 h^{-1}$\Mpc~(Daddi et al.~2002). Such a measurement was 
attempted by Georgakakis et al.~(2000) for sources with $S_{\rm{1.4}}>0.5$\mJy~
in the $\simeq 3$deg$^{2}$ Phoenix radio survey (Hopkins et al.~1998). Whilst 
they found some evidence for clustering, the uncertainties were such that they 
could not determine whether the starbursts have a lower $r_{\rm{0}}$ than the 
radio-loud AGN.

To overcome these limitations we have initiated The Bootes Deep Field, a radio 
survey covering $\simeq$6~deg$^{2}$ down to a limiting $5\sigma$ sensitivity 
of 140\microJy~at 1.4~GHz (de Vries et al. 2002). The source finding algorithm 
identifies 3172 distinct sources, of which $\simeq 10$~per cent are resolved 
by the $13 \times 27$~arcsec beam (the actual number of sources used in our 
angular clustering analysis is smaller than this, due to our use of a subset 
of the data with a slightly higher flux limit, and the use of certain 
algorithms to identify multiple component sources; see section~2). It will be 
complemented by imaging in six optical and near-infrared wavebands (as part of 
the {\em NOAO Deep Wide-Field Survey}), and at longer infrared wavelengths 
with SIRTF. These data will enable morphological classification of the radio 
sources into starburst or AGN, photometric redshift estimation, and an 
examination of the far-infrared:radio correlation for starburst galaxies out 
to cosmological distances (building on work by Garrett~2002). Direct 
measurement of the real space correlation functions of both radio-loud AGN 
and starbursts (and of the cross-correlation between them) will also be 
possible to $z \sim 1$. Furthermore, these measurements will be free from the 
confusion caused by multi-component radio sources in existing catalogues 
without identifications. Until this dataset is assembled, however, we are 
confined to use of the angular correlation function, and it is these 
measurements and their interpretation that we present here. 

\section{ANGULAR CLUSTERING MEASUREMENTS}
Full details of the Bootes survey characteristics and source extraction 
procedure are given in de Vries et al.~(2002). As detailed therein, the survey 
is complete down to a limiting flux density of 0.2\mJy, as determined by the 
point at which the source counts exhibit the first systematic deviation from a 
low-order polynomial fit (see their Fig.~9; N.B. The source counts shown in de 
Vries et al.~are all too low by a factor 2.035 due to a numerical error in the 
construction of this plot). Since extracted sources must have peak fluxes at 
least 5 times higher than the local rms noise level, for the clustering 
analysis we exclude the corners of the survey area where the latter exceeds 
0.04\mJy; the remaining area covers 5.32~deg$^{2}$.

The angular two-point correlation function $\omega(\theta)$ is defined through 
the expression for the probability, $\delta P$,  of finding two sources in 
solid angle elements $\delta\Omega_{\rm{1}}$ and $\delta\Omega_{\rm{2}}$, with 
angular separation $\theta$:

\begin{equation}
\delta P = N^{2} \delta\Omega_{\rm{1}}\delta\Omega_{\rm{2}}(1 + \omega(\theta)),
\end{equation}

where $N$ is the mean areal source density. The function $\omega(\theta)$ is 
therefore a measure of the deviation from a random distribution. It can be 
calculated by comparing the number of source pairs within a given range of 
angular separation with the number of pairs in a large random catalogue 
covering the same area. Of the variety of estimators for $\omega(\theta)$ 
which have been proposed, we follow Rengelink~(1999) and use that due to 
Hamilton~(1993) because of its robustness at large angular scales:

\begin{equation}
\omega(\theta) = 4 \frac{n_{\rm{D}}n_{\rm{R}}}{(n_{\rm{D}}-1)(n_{\rm{R}}-1)}\frac{DD \cdot RR}{DR \cdot DR} - 1
\end{equation}

where $DD$ is the number of data-data pairs within the angular bin centred on 
separation $\theta$, and $RR$ and $DR$ are the numbers of random-random and 
data-random pairs, respectively, within the same separation interval. We use 
a random catalogue containing $n_{\rm{R}}=25000$ sources, vastly exceeding the 
size of the data catalogue ($n_{\rm{D}}$). Errors on the individual 
$\omega(\theta)$ points are computed with the method of Ling, Barrow \& 
Frenk~(1986), by calculating the standard deviation in $\omega(\theta)$ among 
20 pseudo-random resamples of the observational dataset. Such estimates exceed 
the Poisson errors, which are correct only for unclustered data.

Using this procedure, $\omega(\theta)$ was computed in equally-spaced 
logarithmic angular intervals between 0.25 and 155 arcmin for several 
subsamples with lower limiting fluxes between 0.2 and 2\mJy. Each was then 
fitted with a function of the form:

\begin{equation}
\omega(\theta)_{\rm{fit}} = A \theta^{-\delta} - C
\end{equation}

with $\theta$ in degrees. The quantity C (the integral constraint; Groth \& 
Peebles~1977) is a bias resulting from the finite boundary of the survey and 
is given by:

\begin{equation}
C = \frac{1}{\Omega^{2}} \int \int \omega(\theta) d \Omega_{1} d \Omega_{2}.
\end{equation}

For the conventional power-law form $\omega(\theta) = A \theta^{-\delta}$, 
Monte-Carlo computation of the integral yields $C=1.154A$ and $1.392A$ for 
$\delta=0.8$ and $1.1$, respectively.

A complication in the analysis of radio source clustering is the signal at 
small angular separations (below a few arcminutes) caused by sources with 
multiple components. This problem will be largely alleviated when such sources 
can be reliably identified with the optical and infrared data. In the 
meantime, we begin by adopting some prescriptions used by Georgakakis et 
al.~(2000), and before that by Magliocchetti et al.~(1998) and Cress et 
al.~(1996), to identify the genuine multi-component sources prior to measuring 
the angular correlation function. Using the $\theta \propto \sqrt{S}$ relation 
found by Oort~(1987), we consider as a single object all doubles with 
$\theta < 20 \sqrt{F_{\rm{total}}}$, where $\theta$ is their separation in 
arcseconds and $F_{\rm{total}}$ their summed flux density in \mJy; 
furthermore, we only collapse doubles whose individual component fluxes 
differ by less than a factor of 4 (since components of genuine doubles are 
expected to have correlated fluxes; Magliocchetti et al.~1998). Of the 214 
double sources identified by the source extraction algorithm, 50 are 
identified as genuine by these criteria. The 48 sources with 3 or more 
components (39 triples, 9 quadruples) were examined by eye and subjective 
criteria (e.g. relative component fluxes, morphology) were used to decide 
whether or not to treat them as a single source, resulting in the assignment 
of 78 separate sources.

\begin{figure}
\includegraphics[width=0.46\textwidth,angle=0]{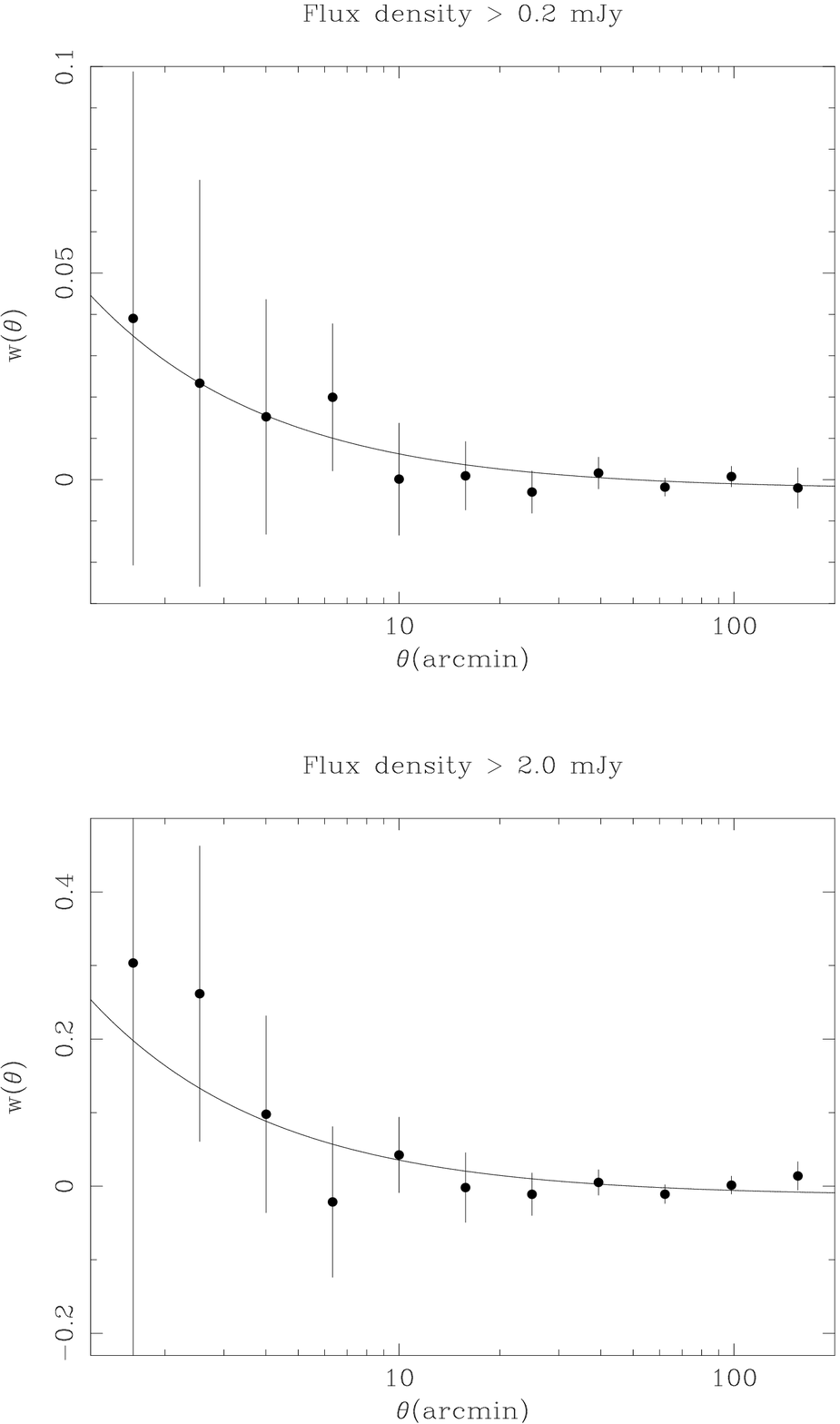}
\caption{\normalsize Angular correlation functions for subsamples with lower 
flux limits of 0.2 and 2\mJy, after the attempted removal of genuine 
multi-component radio sources, as described in section 2. The solid lines show 
fits to the data over $1.5 \leq \theta \leq 20$~arcmin with the function given 
in eqn.~3, for $\delta=0.8$. There is an excess of data over the model at the 
smallest separations in the 2\mJy~data, likely due to the residual effects of 
the multi-component sources.}
\label{fig:angcorr1}
\end{figure}

\begin{figure}
\includegraphics[width=0.35\textwidth,angle=270]{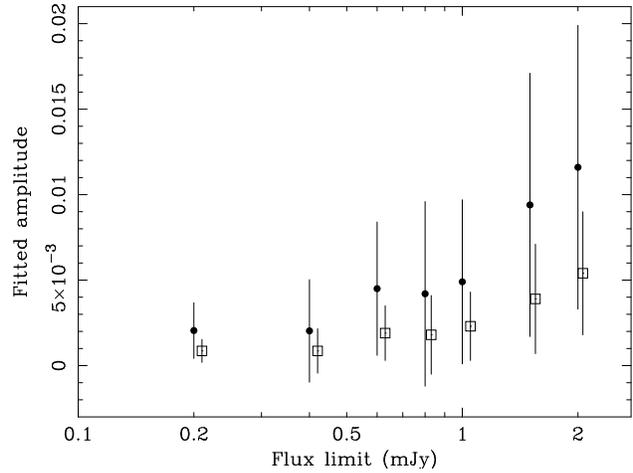}
\caption{\normalsize Variation of the fitted amplitude of the angular 
correlation function with flux limit in The Bootes Deep Field (after attempted 
removal of multi-component sources from the catalogue), for assumed power-law 
indices of $\delta=0.8$ (filled circles) and $\delta=1.1$ (open squares); for 
clarity the latter points have been displaced slightly in flux.}
\label{fig:ampfluxBootes}
\end{figure}

Fits to the data were performed between separations of 1.5 and 20 arcmins, and 
are shown in Fig.~\ref{fig:angcorr1} for the 0.2 and 2\mJy~subsamples. The 
fitted parameters for these and other subsamples are listed in Table~1. As 
expected, the amplitudes are lower when fitted with steeper power-laws. The 
variation of amplitude with flux limit for our survey is shown in 
Fig.~\ref{fig:ampfluxBootes}. Although the error bars are large, our measured 
amplitudes of $\simeq 0.01$ for flux limits of 1--2\mJy~are consistent with 
the results of Georgakakis et al~(2000) for the Phoenix survey, and with those 
of Cress et al.~(1996) who measured an amplitude of $\simeq 0.008$ for FIRST 
sources in the flux interval 1--2\mJy.

However, clustering analyses of the NVSS and FIRST radio surveys by BW02 and 
Overzier et al.~(2002) are at odds with the earlier measurements from the 
FIRST survey. Both papers question the efficacy of the procedures previously 
used (which we also adopt) to identify the multi-component sources: they find 
that the angular correlation function below 6 arcmin is dominated by 
multi-component sources and that the true cosmological clustering amplitude is 
essentially constant at $\simeq 10^{-3}$ from 3\mJy~to $\simeq 50$\mJy. It 
seems implausible that the amplitude could drop by a factor of 10 from flux 
limits of 1--2\mJy~to 3\mJy. We thus conclude that the increase in amplitude 
from 0.2 to 2\mJy~in Fig.~\ref{fig:ampfluxBootes} is due to the residual 
effects of multi-component sources at the higher fluxes. Indeed, 
Fig.~\ref{fig:angcorr1} shows excess signal above the fitted power-law at 
small angles for the 2\mJy~subsample. 


\begin{table}
\caption{Fitted amplitudes of the angular correlation function after attempted 
removal of multi-component sources. The errors are $1 \sigma$.}
\begin{tabular}{|lll|} \hline
Flux density (\mJy) & $10^{3}A(\delta=0.8)$ & $10^{3}A(\delta=1.1)$ \\ \hline
$>0.2$	& 2.1 $\pm$ 1.6 	&  0.86 $\pm$ 0.67 \\
$>0.4$	&  2.0 $\pm$ 3.0  & 0.86 $\pm$ 1.3 \\
$>0.6$	& 4.5 $\pm$ 3.9 & 1.9 $\pm$ 1.6 \\
$>0.8$	& 4.2 $\pm$ 5.4 &  1.8 $\pm$ 2.3 \\
$>1.0$	&  4.9 $\pm$ 4.8	& 2.3 $\pm$ 2.0 \\
$>1.5$	& 9.4 $\pm$ 7.7	& 3.9 $\pm$ 3.2 \\
$>2.0$	& 12 $\pm$ 8.3 & 5.4 $\pm$ 3.6 \\ \hline
\end{tabular}
\end{table}

\begin{figure}
\includegraphics[width=0.46\textwidth,angle=0]{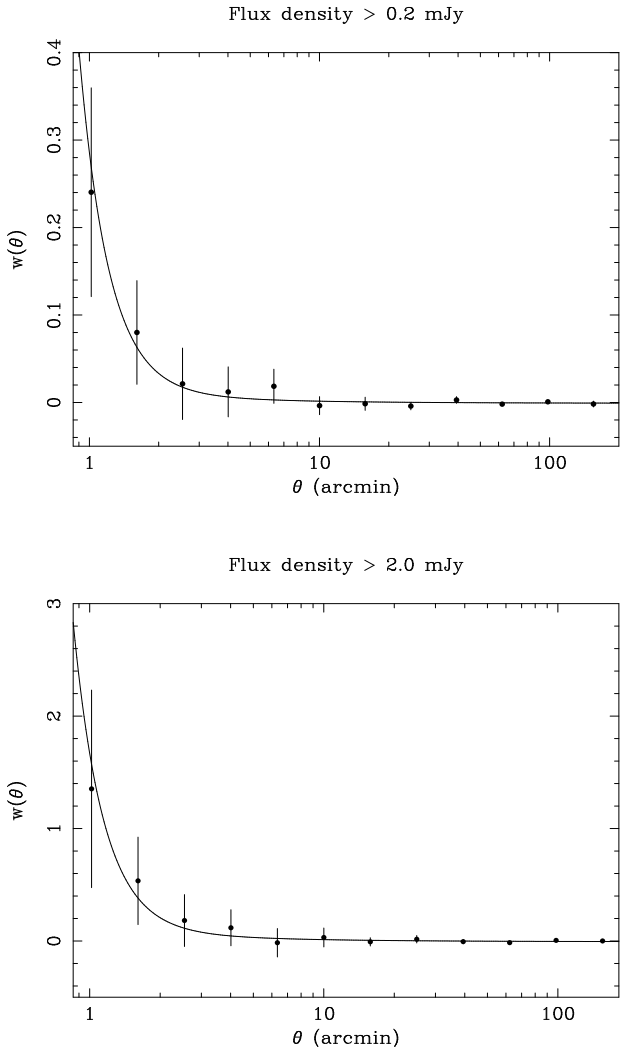}
\caption{\normalsize Angular correlation functions for subsamples with lower 
flux limits of 0.2 and 2\mJy, {\em without} applying any component combining 
procedures to the catalogue; the solid lines are fits to the data above 1 
arcmin with a double power-law, as described in the the text. Note the 
different y-axis scales.}
\label{fig:angcorr2}
\end{figure}

\begin{figure}
\includegraphics[width=0.35\textwidth,angle=270]{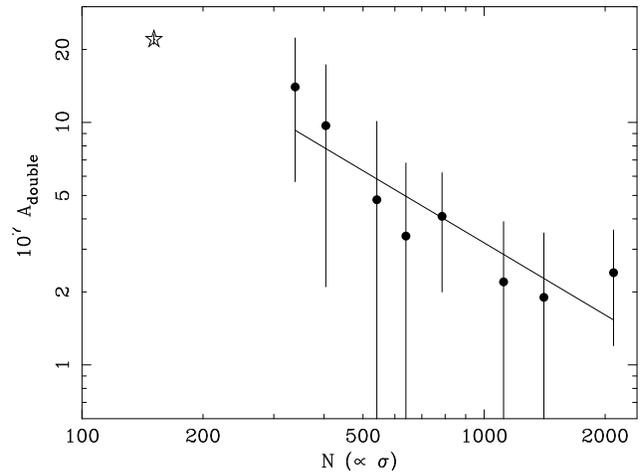}
\caption{\normalsize The circles represent the amplitude of the contribution 
to $\omega(\theta)$ from the multi-component sources, from Table 2. $N$ is the 
number of sources above the corresponding flux limit ($\propto \sigma$, the 
surface density -- with no component-combining procedures applied). The fit to 
these points with $A_{\rm{double}} \propto N^{\alpha}$ is shown (with 
$\alpha=-0.99 \pm 0.6$), and is consistent with the $N^{-1}$ extrapolation 
from the higher flux limits found by BW02 (whose point for sources $>5$\mJy~is 
indicated by the star).}
\label{fig:doubleamp}
\end{figure}

In an attempt to overcome the effects of multi-component sources, we instead 
apply none of the above component combining procedures and fit the resulting 
correlation functions with the sum of two power-laws of the form shown in 
eqn~(3). The first has $\delta=3.4$ (fixed) and is due to the multi-component 
sources which dominate the signal below 6~arcmin, as found by BW02 and 
Overzier et al. (the corresponding integral constraint is $A=1191C$, with the 
correlation function truncated below the angular resolution of the survey to 
keep the integral in eqn.~4 finite). The second has $\delta=0.8$ (fixed) and 
is due to the cosmological clustering. Fits are performed to data above 
1~arcmin, as shown in Fig.~\ref{fig:angcorr2}~for the 0.2 and 2\mJy~sub-
samples. Table 2 lists the results. There is now no significant detection of 
cosmological clustering and the effects of the multi-component sources 
dominate the $\omega(\theta)$ signal. Furthermore, Fig.~\ref{fig:doubleamp} 
shows that the amplitude of the size distribution power-law is consistent with 
the extrapolation to fainter fluxes of the $1/\sigma$ dependence ($\sigma$ 
being the surface density of radio sources) found from 5 to 50\mJy~by BW02. 
With reference to their section 3, this is consistent with the quantity 
$e/\bar{n}$ varying by no more than a factor of $\simeq 2$ from 50 to 0.2\mJy, 
as judged from the scatter around the fit in Fig.~\ref{fig:doubleamp} ($e$ 
being the fraction of sources observed to have multiple components, $\bar{n}$ 
the average number of radio components per source).

\begin{table}
\caption{Fitted amplitudes of the size-distribution and cosmological 
clustering power-laws derived from a two component fit. The errors are 
$1 \sigma$.}
\begin{tabular}{|lll|} \hline
Flux density & size power-law & cosmological clustering \\
(\mJy)       & $10^{7}A(\delta=3.4)$ & $10^{3}A(\delta=0.8)$ \\ \hline
$>0.2$	& 2.4 $\pm$ 1.2 & 0.55 $\pm$ 1.4 \\
$>0.3$  & 1.9 $\pm$ 1.6 & 0.38 $\pm$ 1.9 \\
$>0.4$	& 2.2 $\pm$      1.7 & 1.2  $\pm$      2.5  \\
$>0.6$	& 4.1  $\pm$     2.1 & 1.5  $\pm$      3.7  \\
$>0.8$	& 3.4  $\pm$     3.4 & 2.1  $\pm$      4.4   \\
$>1.0$	&  4.8 $\pm$     5.3 & 2.5  $\pm$     4.7   \\
$>1.5$	& 9.7  $\pm$     7.6 & 4.8  $\pm$     6.9  \\
$>2.0$	& 14  $\pm$     8.3 & 4.3  $\pm$      8.5 \\ \hline
\end{tabular}
\end{table}

\section{INTERPRETATION}
We now model the evolution of the space density and clustering of the sub-\mJy~
radio source population, to determine in particular the effect on the 
clustering signal of the appearance of the starburst galaxies. The spatial and 
angular correlation functions, $\xi(r)$ and $\omega(\theta)$, are related via 
the relativistic Limber equation (see e.g. Peebles 1980; Magliocchetti et al.~
1999); when $\xi(r,z)=(r/r_{\rm{0}})^{-\gamma} (1+z)^{-(3+\epsilon)}$ ($r$ and 
$r_{\rm{0}}$ being proper lengths), $w(\theta)=A \theta^{-(\gamma - 1)}$ 
($\theta$ in radians) where for a spatially flat cosmology with non-zero 
cosmological constant ($\Omega_{\rm{\Lambda}} + \Omega_{\rm{M}} = 1$):

\begin{equation}
A = B \frac{\int_0^{\infty} N^{2}(z) (1+z)^{\gamma - 3 - \epsilon} x^{1 - \gamma} Q(z) dz}{\left(\int_0^{\infty} N(z) dz \right)^{2}}
\end{equation}

with

\begin{equation}
B= \sqrt{\Omega_{\rm{M}}} \left(\frac{r_{\rm{0}} H_{\rm{0}}}{c}\right)^{\gamma} \frac{\Gamma(\frac{1}{2}) \Gamma(\frac{\gamma}{2})}{\Gamma(\frac{\gamma-1}{2})},
\end{equation}

\begin{equation}
Q(z) = \left[(1+z)^3 + \Omega_{\rm{M}}^{-1} -1 \right]^{0.5},
\end{equation}

\begin{equation}
x = \frac{1}{\sqrt{\Omega_{\rm{M}}}} \int_0^{z} \frac{dz}{Q(z)},
\end{equation}

and $N(z)$ is the redshift distribution of sources above the flux limit. When 
a sample comprises two sub-populations, A and B (in this case AGN and 
starbursts), with different clustering properties, the signal for the whole 
is given by:

\begin{equation}
\omega_{\rm{eff}} = f_{\rm{A}}^2\omega_{\rm{A}} + f_{\rm{B}}^2\omega_{\rm{B}} +
2f_{\rm{A}}f_{\rm{B}}\omega_{\rm{AB}},
\end{equation}

where $f_{\rm{A}}$ and $f_{\rm{B}}$ are the fractions of the population in the 
two classes, and $\omega_{\rm{A}}$ and $\omega_{\rm{B}}$ their individual 
correlation functions. The term $\omega_{\rm{AB}}$ is the cross-correlation 
between the two populations. In our analysis, we use the Limber equation to 
compute separate $\omega(\theta)$ for the AGN and starbursts, and then combine 
them using eqn.~(9). For simplicity, we set the cross-correlation to zero. 

\begin{figure}
\includegraphics[width=0.35\textwidth,angle=270]{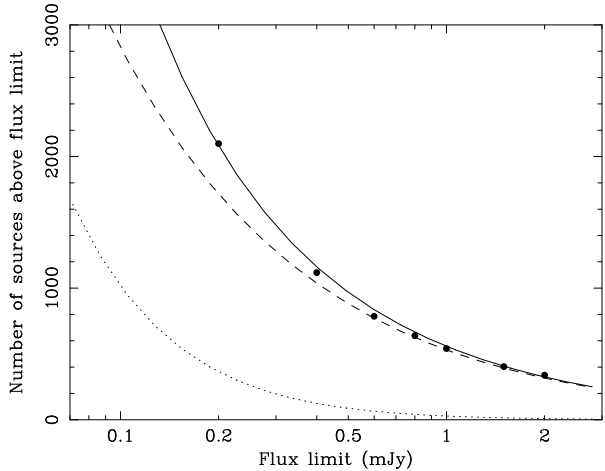}
\caption{\normalsize The points show the integral source counts in the central 
5.32 deg$^{2}$ of The Bootes Deep Field where the catalogue is complete down to
0.2\mJy. The dashed and dotted lines show the contributions of the AGN and the 
starbursts, respectively, and the solid line their sum. The $\sqrt{N}$ error 
bars are smaller than the symbols. See the text for details.}
\label{fig:counts}
\end{figure}

\begin{figure}
\includegraphics[width=0.35\textwidth,angle=270]{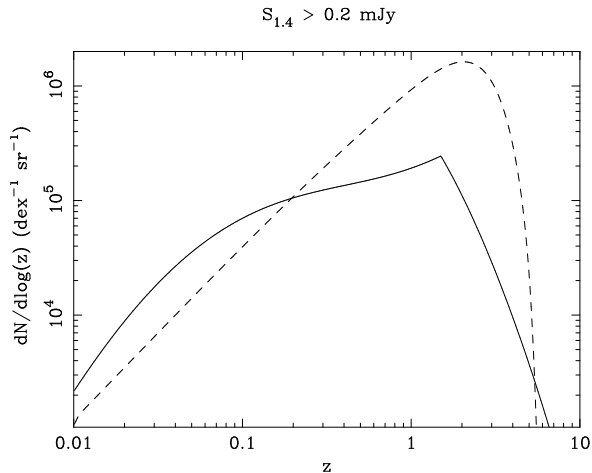}
\caption{\normalsize Model redshift distribution for flux limit of 0.2\mJy. 
The solid and dashed lines denote the starbursts and AGN, respectively.}
\label{fig:zdist}
\end{figure}

Redshift distributions for each flux-limited subsample were computed using the 
Dunlop and Peacock~(1990) pure luminosity evolution model for the combined 
population of flat ($\alpha=0$; $S_{\rm{\nu}} \propto \nu^{\alpha}$) and 
steep ($\alpha=-0.8$) spectrum AGN (taking the parameters from their table C3, 
shifted to 1.4~GHz). For the star-forming galaxies, we use the determination 
of their local 1.4~GHz luminosity function given by Sadler et al.~(2002), with 
pure luminosity evolution of the form $(1+z)^{Q}$ out to $z=z_{\rm{cut}}$, 
with no further evolution thereafter (RR93); with $Q=3.1$ (as in RR93), we 
find that $z_{\rm{cut}}=1.5$ can reproduce the integral source counts 
satisfactorily, as shown in Fig.~\ref{fig:counts} (we plot integral, rather 
than differential, source counts as the former more clearly show the number 
of sources within each flux-limited sample used for the clustering analysis). 
For flux thresholds varying from 0.2 to 2.0\mJy, the model predicts that the 
starburst fraction in the integral source counts decreases from around 0.2 to 
0.03. Fig.~\ref{fig:zdist} shows the model redshift distributions for a flux 
limit of 0.2\mJy. It should be noted that the source counts (in 
Fig.~\ref{fig:counts}) and redshift distribution (in Fig.~\ref{fig:zdist}, and 
used as an input to eqn.~5), were obtained with a cosmology of 
$H_{\rm{0}}=50$\kmpspMpc~and $\Omega_{\rm{M}}=1.0$, for which the above 
luminosity functions were also derived. However, redshift distributions and 
source counts are observational quantities, and thus {\em independent} of the 
assumed cosmology. This is not, therefore, inconsistent with our use of a 
different cosmology ($\Omega_{\rm{M}}=0.3$, $\Omega_{\rm{\Lambda}}=0.7$) in 
the clustering calculations which follow.

\begin{figure}
\includegraphics[width=0.35\textwidth,angle=270]{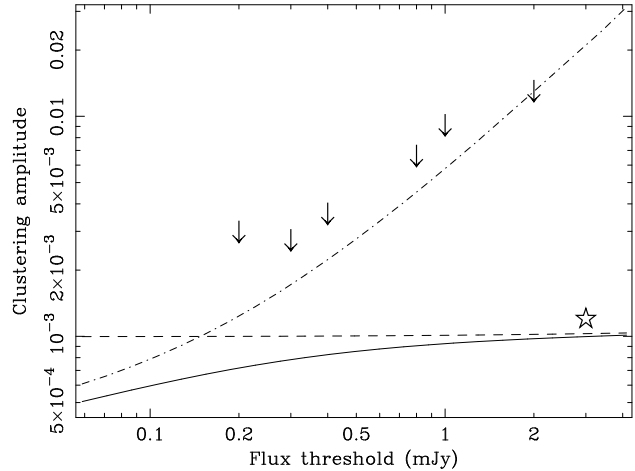}
\caption{\normalsize The solid line shows the model-predicted amplitude of the 
angular correlation function as a function of flux limit. The predicted 
contributions of the AGN and starbursts {\em alone} (i.e. not weighted by the 
squares of the population fractions as shown in eqn.9), are denoted by the 
dashed and dot-dashed lines, respectively. The arrows denote the 90 per cent 
confidence upper limits on the galaxy-galaxy clustering amplitude; the star is 
the measurement from Overzier et al.~(2002) using FIRST and NVSS data.}
\label{fig:modelamp}
\end{figure}

Using the above formalism, we compute the amplitude of the total angular 
correlation function for a baseline model in which 
$r_{\rm{0}}=6.0 h^{-1}$\Mpc~for the AGN (comparable to that measured by BW02), 
and $r_{\rm{0}}=3.0 h^{-1}$\Mpc~for the starbursts (within the range measured 
for local IRAS starburst galaxies). For both correlation functions, we take 
$\gamma=1.8$ and the clustering evolution parameter $\epsilon = \gamma - 3$; 
the latter corresponds to constant clustering in co-moving coordinates, as 
observed for elliptical galaxies (see section 1). We take 
$\Omega_{\rm{M}}=0.3$ and $\Omega_{\rm{\Lambda}}=0.7$ and note that the 
results are independent of $H_{\rm{0}}$. In Fig.~\ref{fig:modelamp} we show, 
as functions of the flux limit, the amplitudes of the angular correlation 
functions of the AGN and starbursts separately, and for the combined 
population (combining the two signals as in eqn.~9). The upper limits in this 
figure are our 90 per cent confidence limits on the amplitude of the 
galaxy-galaxy clustering, obtained from a two power-law fit {\em with the 
amplitude of the size distribution power-law fixed at the value set by the 
$1/\sigma$ extrapolation of the 5\mJy~point of BW02 
(see Fig.~\ref{fig:doubleamp}).} The constancy of the model AGN clustering 
amplitude reflects the invariant shape of the AGN redshift distribution over 
this range in flux limit. The large amplitude of the starburst clustering at 
the high flux end reflects the fact that the brightest starbursts are local 
objects with a fairly narrow redshift distribution; moving towards fainter 
fluxes the redshift distribution broadens, reducing the amplitude. Given the 
small starburst fraction in the model over this flux range, the effective 
signal for the combined population is, with reference to eqn.~(9), essentially 
$\omega_{\rm{eff}} \simeq f_{\rm{AGN}}^{2} \omega_{\rm{AGN}}$. There is thus 
little scope for measuring the clustering of starburst galaxies alone with our 
present data. In order to do so using unidentified radio data one would have 
to push down well below 0.1\mJy~to the level at which the starburst fraction 
becomes substantially higher. However, when the $\sim400$ IRAS-type starburst 
galaxies above the 0.2\mJy~flux limit (see Fig.~\ref{fig:counts}) can be 
identified in the follow-up data due on this field, measurement of their 
clustering strength will become easier.

\section{CONCLUSIONS}
The Bootes Deep Field is a $2.5 \times 2.5$~deg region which has been surveyed 
at 1.4~GHz down to an rms noise level of 28~$\mu$Jy~at its centre. The area 
will be covered at 325~MHz, in six optical and near-infrared bands, and at 
longer infrared wavelengths as part of a SIRTF legacy programme. Here we have 
measured the angular two point correlation function of the 1.4~GHz sources 
down to the survey limit of 0.2\mJy. We find that the size distribution of 
multi-component radio galaxies dominates the overall signal even at these 
faint fluxes, with an amplitude consistent with the extrapolation of the 
$1/\sigma$ variation established over flux limits from 5 to 50 \mJy~by BW02 
($\sigma$ being the surface density of radio sources). This implies that the 
fraction of multi-component sources (normalised by the average number of 
components per source) varies by no more than a factor of 2 from 50\mJy~to 
0.2\mJy.

Only upper limits can be placed on the strength of any true galaxy-galaxy 
clustering. These limits are, however, consistent with the extrapolation of 
the amplitudes measured at higher fluxes by e.g. BW02, and with a model in 
which the clustering of radio-loud AGN is effectively `diluted' by the more 
weakly clustered starburst galaxies. Source count models imply the latter 
population comprises 20 per cent of the sources above 0.2\mJy~(some 400 
objects), with a broad redshift distribution peaking at $z \sim 1-2$. 
Measurement of the clustering of these galaxies alone would extend earlier 
IRAS measurements from $z \sim 0.2$, but must wait until they can be 
identified with the follow-up data on this field.

\section*{ACKNOWLEDGMENTS}
We thank Chris Blake and Emanuele Daddi for discussions and the referees for constructive reports. The second referee is particularly thanked for bringing to our attention the numerical error in the construction of Fig.~9 of de Vries et al.~(2002). RJW acknowledges support from an EU Marie Curie Fellowship.

{}

\end{document}